\documentclass[conference]{IEEEtran}

\usepackage{graphicx}
\usepackage{epstopdf}
\usepackage{standalone}
\usepackage[nolist ]{acronym}
\usepackage{tcolorbox}

\usepackage{pdfpages}
\usepackage{tikz}
\usetikzlibrary{positioning}
\usepackage[bookmarks=false]{hyperref}
\usepackage{pdfcomment}
\usepackage{hologo}

\usepackage{xstring}

\usepackage[utf8]{inputenc}
\usepackage{marvosym}

\usepackage{algorithm}
\usepackage{algpseudocode}
\usepackage{tikz}

\usepackage{listings}
\definecolor{ListingBackground}{rgb}{0.97,0.97,0.97}
\usepackage{pgfplots}
\pgfplotsset{compat=newest}
\pgfplotsset{
    box plot/.style={
        /pgfplots/.cd,
        fill=blue!30,
        only marks,
        mark=-,
        mark size=0.2em,
        /pgfplots/error bars/.cd,
        y dir=plus,
        y explicit,
    },
    box plot box/.style={
        /pgfplots/error bars/draw error bar/.code 2 args={%
            \draw  ##1 -- ++(.2em,0pt) |- ##2 -- ++(-.2em,0pt) |- ##1 -- cycle;
        },
        /pgfplots/table/.cd,
        y index=2,
        y error expr={\thisrowno{3}-\thisrowno{2}},
        /pgfplots/box plot
    },
    box plot top whisker/.style={
        /pgfplots/error bars/draw error bar/.code 2 args={%
            \pgfkeysgetvalue{/pgfplots/error bars/error mark}%
            {\pgfplotserrorbarsmark}%
            \pgfkeysgetvalue{/pgfplots/error bars/error mark options}%
            {\pgfplotserrorbarsmarkopts}%
            \path ##1 -- ##2;
        },
        /pgfplots/table/.cd,
        y index=4,
        y error expr={\thisrowno{2}-\thisrowno{4}},
        /pgfplots/box plot
    },
    box plot bottom whisker/.style={
        /pgfplots/error bars/draw error bar/.code 2 args={%
            \pgfkeysgetvalue{/pgfplots/error bars/error mark}%
            {\pgfplotserrorbarsmark}%
            \pgfkeysgetvalue{/pgfplots/error bars/error mark options}%
            {\pgfplotserrorbarsmarkopts}%
            \path ##1 -- ##2;
        },
        /pgfplots/table/.cd,
        y index=5,
        y error expr={\thisrowno{3}-\thisrowno{5}},
        /pgfplots/box plot
    },
    box plot median/.style={
        /pgfplots/box plot
    },
    boxplot/every median/.style={
    	ultra thick,dashed,cyan
    }
}

\definecolor{flexicolor}{RGB}{46,49,146}
\definecolor{amaricolor}{RGB}{237,28,36}

\usepackage{xspace}

\usepackage[binary-units=true]{siunitx}
\usepackage{psfrag}
\usepackage{graphicx}
\usepackage{tabularx,booktabs}
\usepackage{multirow}
\usepackage{rotating}

\renewcommand{\baselinestretch}{1}

\usepackage{multirow}
\usepackage[cmex10]{amsmath}
\usepackage[caption=false,font=footnotesize]{subfig}
\usepackage{stfloats}

\newif\ifdoubleblind 

\renewcommand{\baselinestretch}{0.985}

\begin{document}

\newcommand{\paperTitle}{A Reinforcement Learning Approach for Efficient Opportunistic Vehicle-to-Cloud Data Transfer}

\newcommand{\paperAuthors}{Benjamin Sliwa and Christian Wietfeld}
\newcommand{\paperEmails}{$\{$Benjamin.Sliwa, Christian.Wietfeld$\}$@tu-dortmund.de}
\newcommand{\argmax}{\mathop{\mathrm{argmax}}\limits}   

\newcommand{\figurePadding}{0pt}
\newcommand{\figureTopPadding}{\figurePadding}
\newcommand{\figureBottomPadding}{\figurePadding}
\newcommand{\sfw}{0.32}
\newcommand{\sfwDual}{0.48}
\newcommand{\redBox}[1]{\colorbox{red}{#1}\xspace}
\newcommand{\mnoA}{\emph{\ac{MNO}~A}\xspace}
\newcommand{\mnoB}{\emph{\ac{MNO}~B}\xspace}
\newcommand{\mnoC}{\emph{\ac{MNO}~C}\xspace}

\newcommand\q{Q(\mathbf{C_{t}},a)}
\newcommand\qq{Q(\mathbf{C_{t+1}},a)}

\newcommand{\dummy}[3]
{
	\begin{figure}[b!]  
		\begin{tikzpicture}
		\node[draw,minimum height=6cm,minimum width=\columnwidth]{\LARGE #1};
		\end{tikzpicture}
		\caption{#2}
		\label{#3}
	\end{figure}
}

\newcommand{\wDummy}[3]
{
	\begin{figure*}[b!]  
		\begin{tikzpicture}
		\node[draw,minimum height=6cm,minimum width=\textwidth]{\LARGE #1};
		\end{tikzpicture}
		\caption{#2}
		\label{#3}
	\end{figure*}
}

\newcommand{\basicFig}[7]
{
	\begin{figure}[#1]  	
		\vspace{#6}
		\centering		  
		\includegraphics[width=#7\columnwidth]{#2}
		\caption{#3}
		\label{#4}
		\vspace{#5}	
	\end{figure}
}
\newcommand{\fig}[4]{\basicFig{#1}{#2}{#3}{#4}{0cm}{0cm}{1}}

\newcommand{\subfig}[3]
{%
	\subfloat[#3]%
	{%
		\includegraphics[width=#2\textwidth]{#1}%
	}%
	\hfill%
}

\newcommand\circled[1] 
{
	\tikz[baseline=(char.base)]
	{
		\node[shape=circle,draw,inner sep=1pt] (char) {#1};
	}\xspace
}
\begin{acronym}
	\acro{AI}{Artificial Intelligence}
	\acro{WEKA}{Waikato Environment for Knowledge Analysis}
	\acro{DDNS}{Data-driven Network Simulation}
	\acro{GPR}{Gaussian Process Regression}
	\acro{MDP}{Markov Decision Process}
	\acro{RF}{Random Forest}
	\acro{CART}{Classification and Regression Tree}
	\acro{ANN}{Artificial Neural Network}
	\acro{SVM}{Support Vector Machine}
	\acro{M5}{M5 Regression Tree}
	\acro{M2M}{Machine-to-machine}
	\acro{H2H}{Human-to-human}
	\acro{AoI}{Age of Information}
	\acro{NWDAF}{Network Data Analytics Functions}
	\acro{mMTC}{massive Machine-type Communication}
	\acro{RAT}{Radio Access Technology}
	\acro{DRL}{Deep Reinforcement Learning}
	
	\acro{MAE}{Mean Absolute Error}
	\acro{RMSE}{Root Mean Squared Error}
	
	\acro{CASTLE}{Client-side Adaptive Scheduler That minimizes Load and Energy}	
	\acro{ITS}{Intelligent Transportation System}
	\acro{LTE}{Long Term Evolution}
	\acro{eNB}{evolved Node B}
	\acro{MNO}{Mobile Network Operator}
	\acro{UE}{User Equipment}
	
	\acro{TCP}{Transmission Control Protocol}
	\acro{RSRP}{Reference Signal Received Power}
	\acro{RSRQ}{Reference Signal Received Quality}
	\acro{SINR}{Signal-to-noise-plus-interference Ratio}
	\acro{CQI}{Channel Quality Indicator}
	\acro{TA}{Timing Advance}
	
	\acro{CAT}{Channel-aware Transmission}
	\acro{pCAT}{predictive CAT}
	\acro{ML-CAT}{Machine Learning CAT}
	\acro{ML-pCAT}{Machine Learning pCAT}	
	\acro{RL-CAT}{Reinforcement Learning CAT}
	\acro{RL-pCAT}{Reinforcement Learning pCAT}	
\end{acronym}

\acresetall
\title{\paperTitle}

\ifdoubleblind
\author{\IEEEauthorblockN{\textbf{Anonymous Authors}}
	\IEEEauthorblockA{Anonymous Institutions\\
		e-mail: Anonymous Emails}}
\else
\author{\IEEEauthorblockN{\textbf{\paperAuthors}}
	\IEEEauthorblockA{Communication Networks Institute, TU Dortmund University, 44227 Dortmund, Germany\\
		e-mail: \paperEmails}}
\fi

\maketitle

\begin{tikzpicture}[remember picture, overlay]
\node[below=5mm of current page.north, text width=20cm,font=\sffamily\footnotesize,align=center] {Accepted for presentation in: 2020 IEEE Wireless Communications and Networking Conference (WCNC)\vspace{0.3cm}\\\pdfcomment[color=yellow,icon=Note]{
@InProceedings\{Sliwa2020a,\\
	Author = \{Benjamin Sliwa and Christian Wietfeld\},\\
	Title = \{A Reinforcement Learning Approach for Efficient Opportunistic Vehicle-to-Cloud Data Transfer\},\\
	Booktitle = \{2020 IEEE Wireless Communications and Networking Conference (WCNC)\},\\
	Year = \{2020\},\\
	Address = \{Seoul, South Korea\},\\
	Month = \{Apr\},\\
\\
\}
}};
\node[above=5mm of current page.south, text width=15cm,font=\sffamily\footnotesize] {2020~IEEE. Personal use of this material is permitted. Permission from IEEE must be obtained for all other uses, including reprinting/republishing this material for advertising or promotional purposes, collecting new collected works for resale or redistribution to servers or lists, or reuse of any copyrighted component of this work in other works.};
\end{tikzpicture}
\begin{abstract}

%
%
Vehicular crowdsensing is anticipated to become a key catalyst for data-driven optimization in the \ac{ITS} domain. Yet, the expected growth in \ac{mMTC} caused by vehicle-to-cloud transmissions will confront the cellular network infrastructure with great capacity-related challenges.
A cognitive way for achieving relief without introducing additional physical infrastructure is the application of opportunistic data transfer for delay-tolerant applications. Hereby, the clients schedule their data transmissions in a channel-aware manner in order to avoid retransmissions and interference with other cell users.
%
%
In this paper, we introduce a novel approach for this type of resource-aware data transfer which brings together supervised learning for network quality prediction with reinforcement learning-based decision making.
%
%
The performance evaluation is carried out using data-driven network simulation and real world experiments in the public cellular networks of multiple \acp{MNO} in different scenarios. The proposed transmission scheme significantly outperforms state-of-the-art probabilistic approaches in most scenarios and achieves data rate improvements of up to 181\% in uplink and up to 270\% in downlink transmission direction in comparison to conventional periodic data transfer.
\end{abstract}

\IEEEpeerreviewmaketitle

\section{Introduction}

%
%
Within the ongoing transition from human-controlled cars to autonomous traffic systems, the exploitation of the vehicles themselves as moving sensor nodes is a key enabler for data-driven traffic optimization.
%
%
However, while future \acp{ITS} will significantly benefit from high penetration rates of environment-sensing vehicles, the cellular network infrastructure will be confronted with massive increases in resource occupation related to \ac{M2M} communication. A promising approach for improving the resource efficiency of the \emph{existing} network infrastructure is to apply \emph{opportunistic} communication techniques for data-intense \emph{delay-tolerant} applications. In order to avoid wasting cell resources on packet error-related retransmissions, data transmissions are performed in a channel-aware manner with respect to the expected network quality with the goal of minimizing the error probability.

%
%
Similarly, the \emph{anticipatory} communication paradigm \cite{Bui/etal/2017a} proposes to explicitly consider context information (e.g., measurements of the network quality) for optimizing decision processes within communication networks. In previous work, we have presented the client-based opportunistic transmission schemes \ac{CAT}, \ac{pCAT}, and \ac{ML-CAT} \cite{Sliwa/etal/2018b, Sliwa/etal/2019d}, which schedule vehicular sensor data transmissions with respect to the expected network quality. Although those methods are able to achieve massive improvements in the resulting end-to-end data rate and power efficiency, they are a based on heuristics and their large parameter spaces complicate the determination of the optimal operating point.

%
%
\begin{figure}[b]  	
	\vspace{-0.5cm}
	\centering		  
	\includegraphics[width=1\columnwidth]{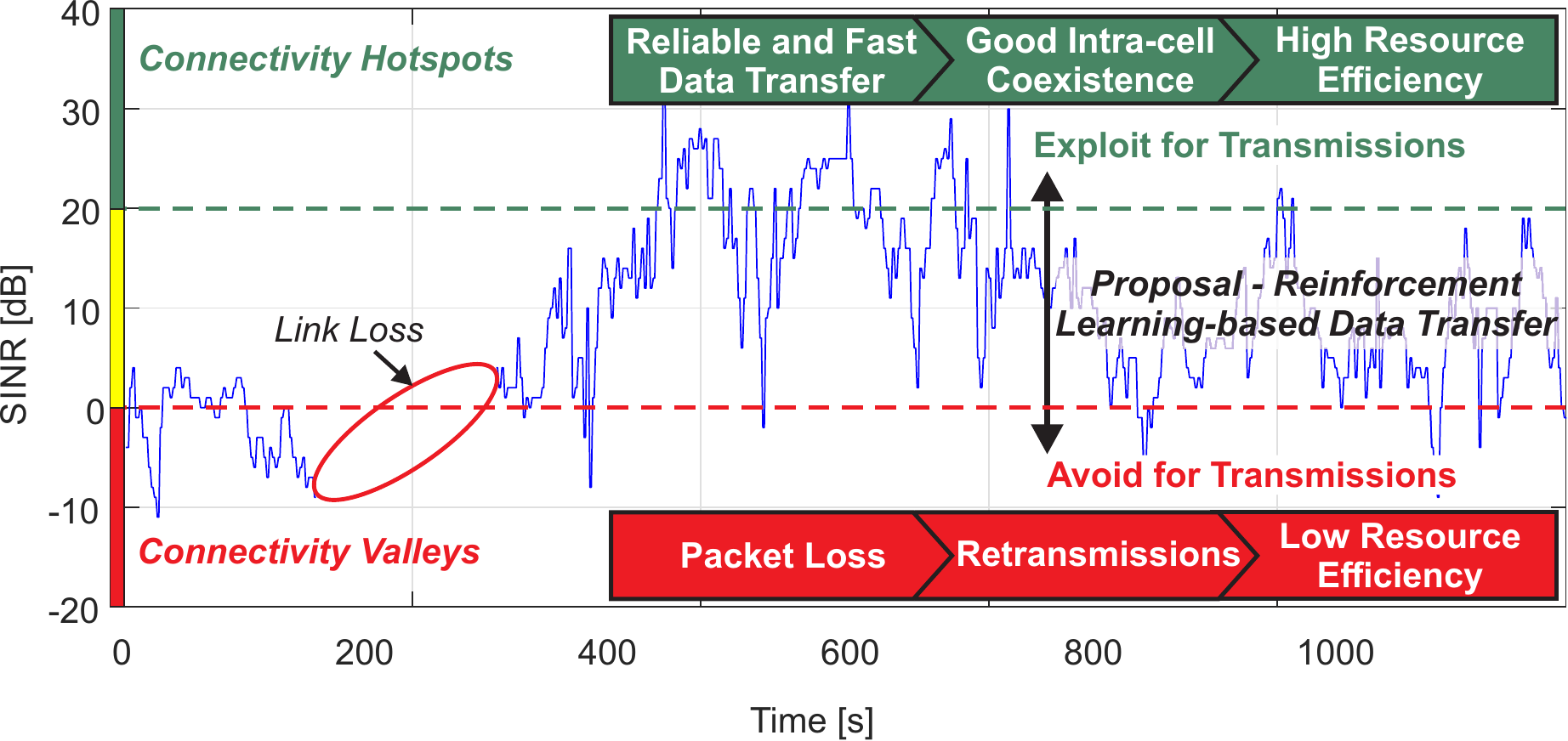}
	\vspace{-0.7cm}
	\caption{Interdependency between network quality and end-to-end transmission behavior. In this paper, reinforcement learning is applied to exploit \emph{connectivity hotspots} and to avoid \emph{connectivity valleys} for reaching the overall goal of improving the resource efficiency of the network.}
	\label{fig:scenario}
\end{figure}
%
%
In this paper, we extend the general ideas of context-aware opportunistic data transfer with a hybrid cognitive networking approach, which brings together \emph{reinforcement learning}-based decision making with supervised machine learning for data rate prediction. The mobile device is modeled as a context-sensing agent, which autonomously learns to detect and exploit favorable transmission opportunities on its own by only considering previously taken actions. For this form of \emph{exploration}, we exploit the findings and the open data set of a large-scale real world data rate prediction campaign, which was presented in \cite{Sliwa/Wietfeld/2019b}.
%
%
Fig.~\ref{fig:scenario} illustrates the real world channel dynamics by means of an \ac{SINR} time series trace and illustrates the involved challenges and opportunities for the vehicular data transfer. Instead of only using \ac{SINR} measurements for channel quality assessment, we jointly consider nine different \ac{LTE} network quality indicators within the proposed reinforcement learning-based approach, which are brought together by a data rate prediction model.

%
%
The contributions provided by this paper are as follows:
\begin{itemize}
	\item Presentation of the novel \textbf{reinforcement learning-based} transmission schemes \ac{RL-CAT} and \ac{RL-pCAT} for optimizing the resource efficiency of cellular vehicular sensor data transmissions.
	\item Proof-of-concept \textbf{real world performance evaluation} and comparison with state-of-the-art probabilistic data transfer approaches.
	\item The measurement and evaluation software\footnote{Source code is available at https://github.com/BenSliwa/MTCApp} as well as the raw results of the performance evaluation \cite{Sliwa/2019a} are provided in an \textbf{open source} way.
\end{itemize}
%
%
%

%
%
The remainder of the paper is structured as follows. In Sec.~\ref{sec:related_work}, we give an overview about data-driven optimization approaches for vehicular data transfer. Afterwards, the proposed reinforcement learning-based transmission schemes are presented in Sec.~\ref{sec:approach}. An overview about the methodological evaluation setup is provided in Sec.~\ref{sec:methods}. Finally, the results of the performance evaluation are presented and discussed in Sec.~\ref{sec:results}.
\section{Related Work} \label{sec:related_work}

%
%
\textbf{Machine learning} is a catalyst for anticipatory communication in complex environments, as it allows to implicitly consider the hidden interdependency between observable measurement variables, which are too complex to bring together in a closed analytical description. An overview about different machine learning approaches for wireless communication systems is given in \cite{Jiang/etal/2017a}. Furthermore, the authors of \cite{Ye/etal/2018a} provide a comprehensive overview about machine learning methods for vehicular communication networks.
%
%
\emph{Supervised} learning models utilize \emph{labeled} data for training of \emph{regression} models, which can then be used for performing \emph{predictions} on unlabeled data. Typically, the models are trained offline and then deployed to the target platform for online application. If major changes of the environment occur -- e.g., a previously unobserved packet scheduler is deployed to the cellular network by the \ac{MNO} -- novel data needs to be obtained and the prediction models need to be re-trained.

%
%
In contrast to that, \emph{reinforcement learning} \cite{Sutton/Barto/2018a} introduces the concept of \emph{cognitive} decision making, where a virtual \emph{agent} senses the environment and optimizes a certain behavior by learning from the \emph{rewards} of taken \emph{actions}. A comprehensive overview about applying this type of machine learning to mobile communication systems is provided by \cite{Gacanin/2019a}.

%
%
\textbf{Data rate prediction} has been proposed as a method for increasing the context-awareness of vehicular communication systems through high-level optimization techniques, e.g., for predictive caching \cite{Mangla/etal/2016a} and multi-\ac{RAT} interface selection \cite{Bouali/etal/2016a}.
Consequently, different research works have investigated client-based data rate prediction in vehicular cellular networks, for which the main findings are summarized as follows:
\begin{itemize}
	\item The radio channel characteristics have a severe impact on the resulting end-to-end data rate. Passively measurable indicators such as \ac{RSRP}, \ac{RSRQ} and \ac{SINR} can provide meaningful information for data rate prediction models \cite{Sliwa/etal/2019d, Sliwa/Wietfeld/2019b, Akselrod/etal/2017a, Riihijarvi/Mahonen/2018a, Jomrich/etal/2018a}.
	\item Integrating the payload size into the prediction process allows to implicitly consider cross layer interdependencies (e.g., the slow start mechanism of \ac{TCP}), which have a strong interdependency with the channel coherence time \cite{Sliwa/etal/2019d, Sliwa/Wietfeld/2019b}.
	\item Crowdsensed connectivity maps can be applied to maintain radio condition data bases, which allow to forecast the network situations vehicles are going to encounter on their future trajectories. Furthermore, the applied cell-wise aggregation implicitly compensates short-term prediction errors (e.g., related to multipath fading) \cite{Sliwa/etal/2019d, Sliwa/Wietfeld/2019b}.
	\item In the vast majority of the evaluations (e.g., \cite{Sliwa/etal/2019d, Sliwa/Wietfeld/2019b, Jomrich/etal/2018a, Samba/etal/2017a}), \ac{CART}-based methods such as \acp{RF} \cite{Breiman/2001a} outperform more complex models (e.g., \emph{deep learning} \cite{LeCun/etal/2015a}), which require a significantly higher amount of training data.
	\item Cellular data rate prediction models are highly \acp{MNO}-dependent due to \acp{MNO}-specific configurations of the network infrastructure \cite{Sliwa/Wietfeld/2019b}.
\end{itemize}
%
%
%

%
%
\textbf{\ac{DDNS}}: In addition to using data rate prediction models for context-aware decision making, the trained models themselves can provide the foundation for simulative optimization of cognitive communication systems using \ac{DDNS} \cite{Sliwa/Wietfeld/2019c}. This approach allows to replay available time series measurements of \emph{passive} network quality indicators to analyze the behavior of novel \emph{active} transmission schemes. The end-to-end behavior is represented by a supervised machine learning model, for which the deviation to the real world measurements is learned by a second \ac{GPR}-based machine learning model. The latter transfers the prediction process from the deterministic to the probabilistic domain and allows to generate synthetic -- yet close to reality -- end-to-end indicator profiles by sampling from distribution of the prediction errors. 
As the real world validation in \cite{Sliwa/Wietfeld/2019c} shows, the achieved results are not only significantly more accurate than conventional system-level network simulations \cite{Cavalcanti/etal/2018a}, the result generation process is also more than an order of magnitude faster, which ultimately allows to perform a deeper exploration of the parameter space within the system optimization phase.

%
%
\textbf{Opportunistic data transfer} is a method for optimizing the resource efficiency of data transmissions for \emph{delay-tolerant} applications by integrating the network quality into the transmission process. 
%
%
Recently, the authors of \cite{Lee/etal/2019a} have presented \ac{CASTLE} as a method for distributed client-side scheduling of coordinated transmissions, which exploits machine learning for channel-sensitive load estimation.
In previous work \cite{Sliwa/etal/2018b, Sliwa/etal/2019d}, we have applied probabilistic methods, where a transmissions probability $p_{\text{TX}}(t)$ is calculated based on measurements of a network quality indicator $\Phi(t)$ with a defined value range $\Phi_{\max}-\Phi_{\min}$, application-specific deadlines $\Delta t_{\min}$ and $\Delta t_{\max}$ and a convergence exponent $\alpha$ as
%
%
%
%
\begin{equation}
	p_{\text{TX}}(t) = \begin{cases}
	0 & \Delta t<\Delta t_{\min} \\ 
	1 & \Delta t>\Delta t_{\max} \\ 
	\left( \frac{\Phi(t)-\Phi_{\min}}{\Phi_{\max}-\Phi_{\min}} \right)^{\alpha} & \text{else}
	\end{cases}
\end{equation}
where $\Delta t$ is the passed time since the last successful transmission has been performed. For the basic transmission scheme \ac{CAT}, the transmission metric $\Phi$ is represented by the measured \ac{SINR}, while \ac{ML-CAT} considers the predicted data rate. 
In this work, the goal is to further optimize the data rate improvement by replacing the probabilistic approach with reinforcement learning-based decision making.

\section{Reinforcement Learning-based Opportunistic Vehicular Data Transfer} \label{sec:approach}

In this section, we present the proposed reinforcement learning-based transmission schemes. The overall goal is to learn a context-aware transmission process, which \emph{exploits} connectivity hotspots and \emph{avoids} data transmissions during connectivity valleys.
%
%
\begin{figure}[]  	
	\centering		  
	\includegraphics[width=1\columnwidth]{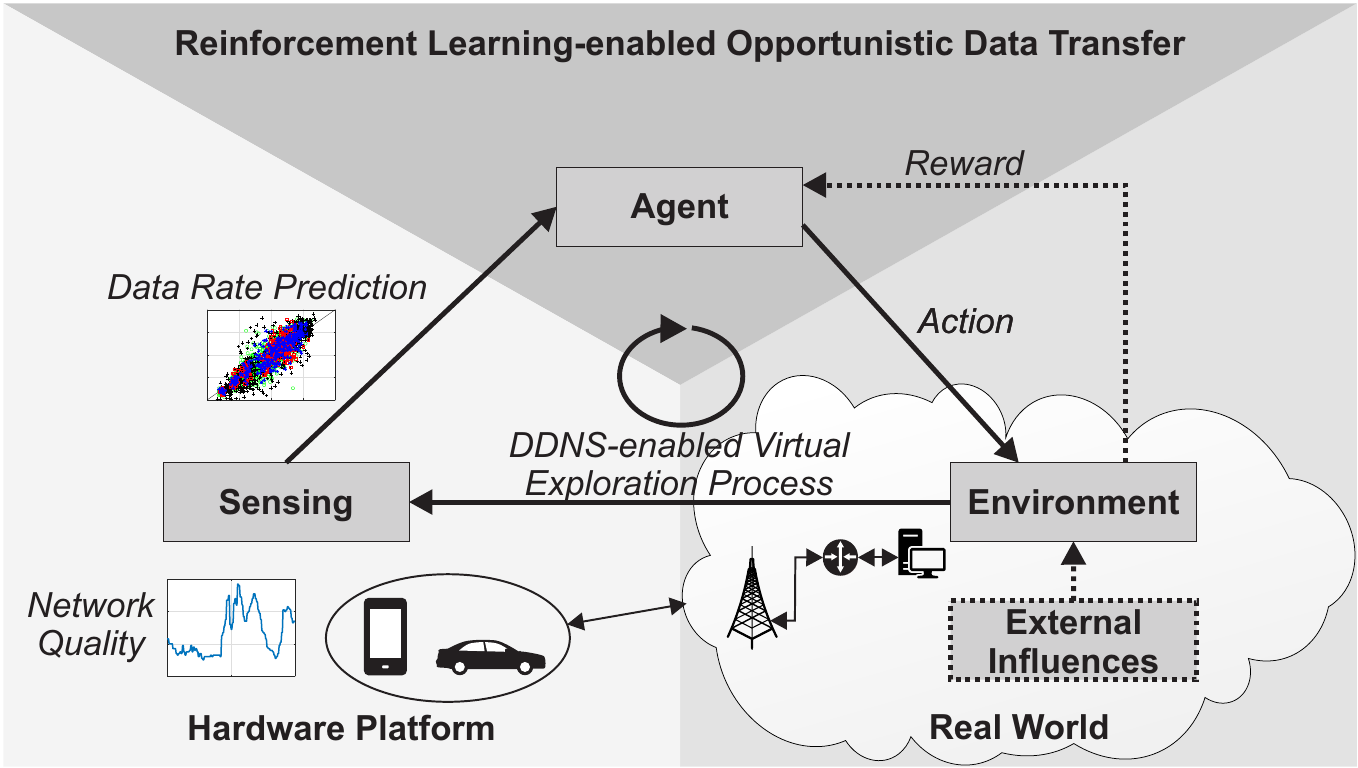}
	\caption{System architecture model: Interaction between agent and environment.}
	\label{fig:reinforcementLearning}
	\vspace{-0.5cm}	
\end{figure}
The overall system architecture model and the interaction between \emph{agent} -- the opportunistic transmission scheme -- and \emph{environment} -- the public cellular network -- is illustrated in Fig.~\ref{fig:reinforcementLearning}.
%
%
The proposed system is composed of three logical domains. The actual decision making is performed in the \emph{agent} domain, where the agents decides if the buffered data shall be transmitted immediately or if the transmission should be postponed as the current network situation is not favorable. The \emph{real world} domain represents the network environment, which mainly impacted by external influence factors (e.g., other cell users, mobility-related channel dynamics) and not by the taken actions of the agent itself. For completeness, it is remarked that transmissions performed by the agent have a minor impact on the environment due to the occupied network resources. The foundation for the decision making is the \emph{sensing} process, which is performed in the \emph{hardware platform} domain. Based on measurements of raw context features, a prediction model is applied to forecast the currently achievable data rate. 
%
%
In this paper, we exploit the high computational efficiency of \ac{DDNS} for implementing a \emph{virtual} exploration process (see Sec.~\ref{sec:exploration}), which trains the agent with data transmission profiles synthesized from previous real world transmissions. 
%
%
Based on this foundation, we derive the reinforcement learning-based opportunistic data transmission scheme \ac{RL-CAT} and its mobility-predictive extension \ac{RL-pCAT} in the next paragraphs.

\subsection{Context-aware Approach: \acf{RL-CAT}} \label{sec:rl_cat}

Reinforcement learning is applied to derive a \emph{decision table} $Q$, which allows to assess the expected rewards by performing the possible actions for a given state. For opportunistic data transfer, the possible actions $a$ are \texttt{IDLE} (data is buffered) and \texttt{TX} (data is transmitted). 
We model the state as a \emph{context tuple} $\mathbf{C}_{t}$ as
%
%
\begin{equation}
	\mathbf{C}_{t} = (\tilde{S}(t), \Delta t)
\end{equation}
with $\tilde{S}(t)$ being the predicted data rate, which is discretized to the closest integer value, and 
$\Delta t$ being the passed time since the last successful transmission has been performed.
As the data rate prediction accuracy is reduced in the edge regions of the data rate value range \cite{Sliwa/Wietfeld/2019b}, the reinforcement learning process immanently learns a confidence model for the machine learning-based data rate prediction.

%
%
\textbf{Online Decision Making:}  
%
%
\fig{b}{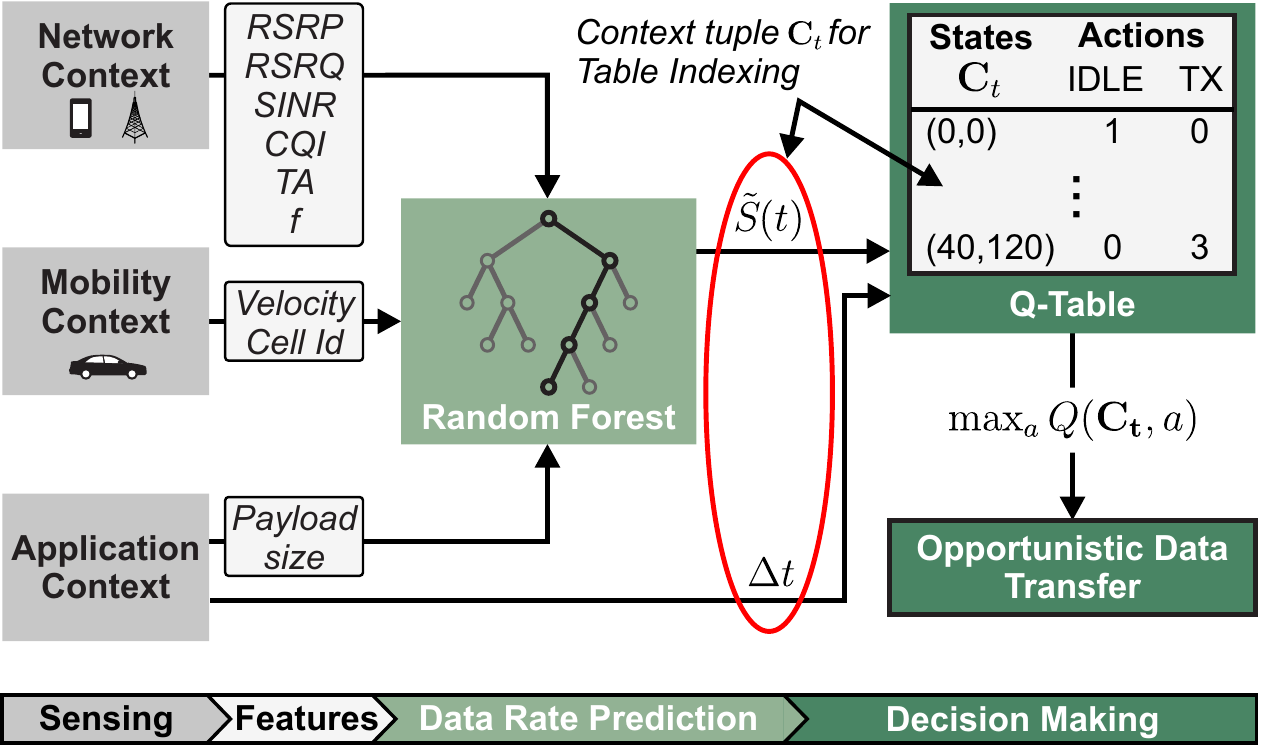}{Online application of the reinforcement learning-based decision making for the basic \ac{RL-CAT} approach. For \ac{RL-pCAT}, the context state is represented by an extended tuple $\mathbf{C}_{t}' = (\tilde{S}(t), \tilde{S}(t+\tau), \Delta t)$.}{fig:features}
An overview of the process for online decision making is illustrated in Fig.~\ref{fig:features}. Different context measurements are used to estimate the currently achievable data rate $\tilde{S}(t)$ with a supervised prediction model. Based on the findings and analyzed model comparisons in \cite{Sliwa/Wietfeld/2019b}, the latter is represented by a \ac{RF} prediction model with maximum depth 15 and 100 trees. The feature vector $\mathbf{F}(t)$ is composed of measurements from different context domains:
%
%
\begin{itemize}
	\item \textbf{Network}: \ac{RSRP}, \ac{RSRQ}, \ac{SINR}, \ac{CQI}, \ac{TA}, Carrier frequency $f$
	\item \textbf{Mobility}: Velocity, Cell id
	\item \textbf{Application}: Payload size of the to be transmitted data packet
\end{itemize}
Using the predicted data rate $\tilde{S}(t)$ within $\mathbf{C_{t}}$ instead of the raw context features significantly reduces the dimension of the table index. Therefore, the exploration phase of the reinforcement learning process can be performed more efficiently, as fewer exploration epochs are required.
%
%
Together with $\Delta t$, the context state $\mathbf{C_{t}}$ is composed and the to be performed action $a$ is then selected by maximizing the achievable $Q$ value with $\max_a \q$.

%
%
\textbf{Iterative Exploration Process}:  
%
%
Before the system is able to make cognitive decisions on its own, it needs to fill the $Q$-table with valid data through an iterative \emph{exploration} process, which considers the rewards of previously performed actions. For the proposed reinforcement learning-based transmission scheme, we apply an adjusted version of the classical \emph{Q-learning} \cite{Watkins/Dayan/1992a} technique. 
%
%
At first, all table entries are randomized with $\mathcal{N}(0, 1)$.
%
%
For each performed action, the $Q$-table is then updated as
%
%
\begin{equation} \label{eq:q_learning}
	\q = (1-\alpha) \cdot \q + \alpha \left[ r_{a} + \lambda  \cdot \max_a \qq \right] 
\end{equation}
with \emph{learning rate} $\alpha$, \emph{reward} $r_{a}$, \emph{discount factor} $\lambda$, and $\mathbf{C_{t+1}}$ being an estimation of the Q-value in the future state after the action $a$ has been taken. 
However, in the considered vehicular scenario, the environment state is changed mainly due to external impact factors -- e.g., mobility-related channel dynamics -- and the taken actions do not have a measurable impact. Therefore, Eq.~\ref{eq:q_learning} is simplified to
%
%
\begin{equation}
	\q = (1-\alpha) \cdot \q + \alpha \cdot r_{a}
\end{equation}
%
%
Separate reward functions $r_{a}$ are applied for the possible actions $a$. The reward $r_{\text{TX}}$ of a performed transmission with measured data rate $S$ is calculated with respect to the trade-off between a defined \ac{MNO}-specific target data rate $S^*$ and an application-specific age of information deadline $\Delta t_{max}$ with a weighting trade-off factor $w$
%
%
\begin{equation}
	r_{\text{TX}} (S, \Delta t) = \frac{w \cdot (S-S^*)}{S_{\max}} + \frac{\Delta t \cdot (1-w)}{\Delta t_{\max}} 
\end{equation}

For \ac{RL-CAT}, the \texttt{IDLE} action is not able to achieve a reward as no data is transferred during the buffering phase. However, a \emph{deadline violation punishment} which is represented by a large negative number $\Omega$ is introduced in order to ensure $Q(\mathbf{C_{t},\texttt{TX}}) >> Q(\mathbf{C_{t},\texttt{IDLE}})$ if the \ac{AoI} deadline is reached, which then causes an immediate data transmission regardless of the expected resource efficiency.
%
%
\begin{equation}
	r_{\text{IDLE}} (\Delta t) = \begin{cases}
	\Omega & \Delta t \geq \Delta t_{\max} \\ 
	0 & \text{else}
	\end{cases}
\end{equation}

\subsection{Context-predictive Approach: \acf{RL-pCAT}} \label{sec:rl_pcat}

%
%
As previous studies \cite{Sliwa/etal/2019d} have shown, opportunistic vehicular data transfer methods can significantly benefit from not only considering the current context, but also taking predictions for the anticipated future context behavior into account. In the following, we therefore extend the basic \ac{RL-CAT} concepts to the context-predictive \ac{RL-pCAT} method which considers the anticipated future network quality along the expected trajectory of the vehicle.
The context tuple is extended by an additional data rate prediction $\tilde{S}(t+\tau)$ for a given temporal look ahead $\tau$ to $\mathbf{C}_{t}'$ as
%
%
\begin{equation}
	\mathbf{C}_{t}' = (\tilde{S}(t), \tilde{S}(t+\tau), \Delta t)
\end{equation}

%
%
However, as the future feature vector $\mathbf{F}(t+\tau)$ cannot be measured at the time of the decision making $t$, it is predicted based on aggregated measurements which were previously performed in the same geographical region. The estimated feature vector $\mathbf{\tilde{F}}(t+\tau)$ is looked up from a multidimensional connectivity map $M$ with cell size $c$ as
\begin{equation}
	\mathbf{\tilde{F}}(t+\tau) = M ( \lfloor \frac{\mathbf{\tilde{P}}(t+\tau)}{c} \rfloor ) 
\end{equation}
%
%
with $\mathbf{\tilde{P}}(t+\tau)$ being an estimation of the future vehicle position, which is derived based on trajectory-aware mobility prediction. Details about the algorithmic implementation and a real world evaluation of the prediction errors as well as their impact on the network quality prediction can be found in \cite{Sliwa/etal/2018a}. 

The reinforcement learning process is performed analogously to Sec.~\ref{sec:rl_cat}. However, the reward function of the \texttt{IDLE} action is changed to $r_{\text{IDLE}}'$ as
%
%
\begin{equation}\label{eq:rl_pcat_idle}
r_{\text{IDLE}}' (\Delta t) = \begin{cases}
\Omega & \Delta t \geq \Delta t_{\max} \\ 
\frac{1}{\tau} \cdot r_{\text{TX}}(\tilde{S}(t+\tau), \Delta t + \tau) & \text{else}
\end{cases}
\end{equation}
as postponing the transmission at $t$ is now immanently related to the predicted context at $t+\tau$.

%
%
It is remarked that the application of context prediction introduces additional error sources to the system which impact the achievable performance. Imperfections of the mobility prediction mechanism might lead to false context lookups and the context aggregation within the connectivity map only represents the mean indicator behavior within the considered cell. In addition, as the data rate prediction is performed for $\tilde{S}(t)$ as well as for $\tilde{S}(t+\tau)$, prediction errors have an increased impact on the channel quality assessment.

\section{Methodology} \label{sec:methods}

In this section, the methods for training the machine learning models and for performing the real world performance evaluation are presented.

\subsection{Machine Learning-enabled Data Rate Prediction} \label{sec:data_rate_prediction}

The training of the data rate prediction model is performed with the \ac{WEKA} \cite{Hall/etal/2009a} framework.
%
%
Based on the findings and the open data sets of \cite{Sliwa/Wietfeld/2019b}, the data rate prediction is performed with a \ac{RF} regression model, which consists of 100 random trees and allows a maximum depth of 15. In order to consider \ac{MNO}-specific characteristics, an individual prediction model is trained for each \ac{MNO} and transmission direction.

%
%
Tab.~\ref{tab:prediction} shows an overview of the coefficient of determination $R^2$, \ac{MAE}, \ac{RMSE} of the \ac{RF}-based data rate prediction in uplink and downlink direction for the three considered \acp{MNO}.
%
%
\newcolumntype{Y}{>{\centering\arraybackslash}X}
\newcommand\mr[1]{\multirow{2}{1cm}{#1}}

\begin{table}[ht]
	\centering
	\caption{Statistical Properties of the Random Forest-based Data Rate Prediction Models}
	
	\begin{tabularx}{\columnwidth}{l c*{6}{Y}}
		\toprule
		\textbf{Model} & \multicolumn{2}{c}{\textbf{\mnoA}} & \multicolumn{2}{c}{\textbf{\mnoB}} & \multicolumn{2}{c}{\textbf{\mnoC}} \\
		\cmidrule(lr){2-3} \cmidrule(l){4-5} \cmidrule(l){6-7}
		& \textbf{UL} & \textbf{DL} & \textbf{UL} & \textbf{DL} & \textbf{UL} & \textbf{DL}\\
		\midrule
				
		%
		%
		\multirow{2}{0.5cm}{$\mathbf{R^2}$} & 0.779 & 0.588 & 0.678 & 0.634 & 0.718 & 0.493 \\
		& $\pm0.023$ & $\pm0.021$ & $\pm0.04$ & $\pm0.062$ &  $\pm0.028$ & $\pm0.036$ \\
		
		%
		%
		\textbf{\acs{MAE}} & 2.984 & 3.302 & 2.603 & 7.01 & 2.537 & 3.136 \\
		\textbf{[MBit/s]} & $\pm0.141$ & $\pm0.113$ & $\pm0.144$ & $\pm0.398$ & $\pm0.117$ & $\pm0.153$ \\

		%
		%
		\textbf{\acs{RMSE}} & 4.061 & 4.743 & 3.619 & 10.177 & 3.424 & 4.276 \\
		\textbf{[MBit/s]} & $\pm0.223$ & $\pm0.21$ & $\pm0.29$ & $\pm1.431$ & $\pm0.168$ & $\pm0.235$ \\		 

		\textbf{Range}  & \mr{39.782} & \mr{42.94} & \mr{38.208} & \mr{159.982} & \mr{35.676} & \mr{33.842} \\
		\textbf{[MBit/s]} & & & & & & \\

		\bottomrule
	\end{tabularx}
	\label{tab:prediction}
	
	\vspace{0.1cm}
	\emph{UL}: Uplink, \emph{DL}: Downlink 
\end{table}

Note that \ac{MAE} and \ac{RMSE} have to be considered with respect to the value range $S_{\max}-S_{\min}$ of the data set. As \mnoB implements downlink carrier aggregation, it achieves a significantly higher value range -- and absolute error measurements -- than the other \acp{MNO}.
%
%
A general observation is that the prediction works better in the uplink than in the downlink transmission direction. As the traffic intensity is typically much higher in the downlink than in the uplink \cite{Bui/etal/2017a}, the resulting downlink data rate is highly impacted by the cell load, which can only be considered indirectly by means of the measurable \ac{RSRQ}. In contrast to that, the uplink performance is more impacted by the network quality dynamics, which are represented by the whole network context feature set utilized by the prediction model.

\subsection{Virtual Exploration Process} \label{sec:exploration}

%
%
Although typical reinforcement learning techniques rely on a controlled trial-and-error mechanism, this method is unfavorable for the considered vehicular scenario as it would require to perform a multitude of real world drive tests in order to reach the convergence level of the proposed transmission schemes (see Sec.~\ref{sec:results_opt}).
To overcome this issue, we apply a \emph{virtual exploration process}, which is modeled within a \ac{DDNS} setup (see Sec.~\ref{sec:related_work} and \cite{Sliwa/Wietfeld/2019c}). 
%
%
\fig{t}{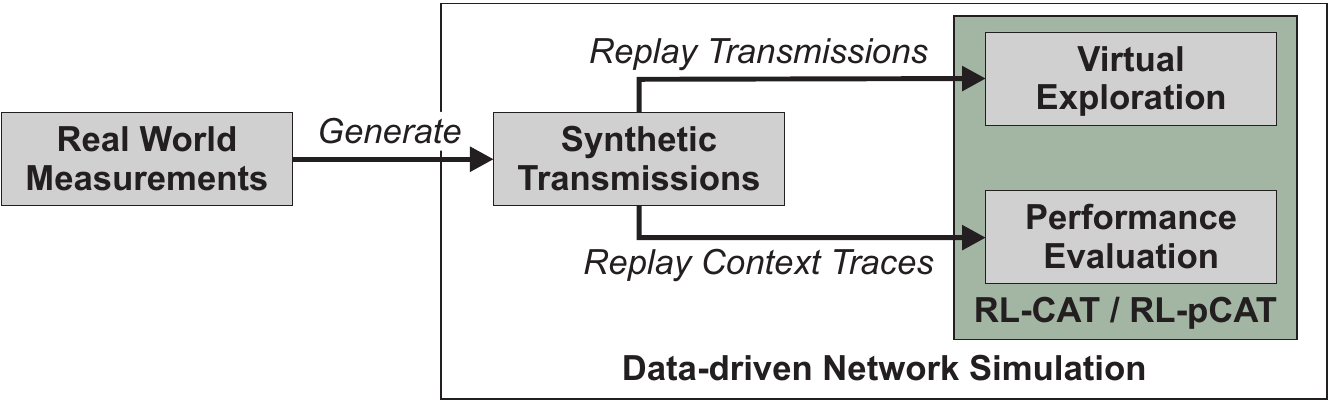}{Overview about the \ac{DDNS}-based virtual exploration process and initial performance evaluation for parameter selection.}{fig:ddns}
%
%
An overview about the involved logical steps is provided in Fig.~\ref{fig:ddns}. Based on the available real world measurements of \cite{Sliwa/Wietfeld/2019b}, we synthesized 2400 network context traces (time series data with two scenarios, periodic transmission interval $\Delta t=\{1,2,...,120~s\}$ with 10 repetitions) for three different \acp{MNO} in uplink and downlink direction. In total, the resulting data set consists of more than 84000 transmissions.
%
%
The generated transmissions and context traces are replayed in random order with the proposed reinforcement learning-based transmission schemes according to Sec.~\ref{sec:rl_cat} and Sec.~ \ref{sec:rl_pcat}, which then learn to perform cognitive data transfer behaviors by identifying favorable and non-favorable transmissions from the previously performed actions. Moreover, the reinforcement learning approach extracts the complex interdependency between network quality and transmission results from the available data sets. 

%
%
For the mobility-predictive \ac{RL-pCAT} transmission scheme, the network context indicators -- which are part of the feature vector of the data rate prediction model -- are aggregated in a multidimensional connectivity map $M$ which is jointly used with trajectory-aware mobility prediction according to \cite{Sliwa/etal/2019d}.
%
%
However, a methodological dilemma needs to be solved as the closed loop scenario only allows to analyze the results of taken actions: If a transmissions is performed at time $t$, the transmission buffer is cleared and it cannot be directly concluded if the same transmission would have achieved a better performance at $t+\tau$. 
Therefore, the another \ac{DDNS} evaluation is carried to analyze the behavior at $t+\tau$ and then update $r_{\text{IDLE}}'$ at time $t$ with Eq.~\ref{eq:rl_pcat_idle}.

\subsection{Real World Performance Evaluation}

After the \ac{DDNS}-based exploration phase, the real world performance evaluation of the converged transmission schemes is carried out in the public cellular \ac{LTE} networks of three different \acp{MNO} in Germany. Data is transmitted in uplink and downlink direction from the vehicle through the cellular network to a cloud-based server. A virtual sensor application generates $50$~KB of sensor data per second which is buffered locally until the whole data buffer is transmitted and cleared. The resulting \ac{AoI} of each successful transmission corresponds to the generation time of the oldest contained sensor packet.
The measurement application is executed on Android-based \acp{UE} (Samsung  Galaxy  S5  Neo,  Model  SM-G903F). We consider two different scenarios (suburban and highway) with different speed characteristics and building densities. For each of the tracks, 10 different drive tests are performed for each of the considered transmission schemes.

A summary about the parameters of the overall system is given in Tab.~\ref{tab:parameters}. Further \ac{MNO}-specific configurations are summarized in Tab.~\ref{tab:mnos}.
%
%
\newcommand{\entry}[2]{#1 & #2 \\}
\newcommand{\head}[2]{\toprule \entry{\textbf{#1}}{\textbf{#2}}\midrule}

\begin{table}[ht]
	\centering
	\caption{Parameters of the Reference Scenario}
	\begin{tabular}{ll}
		
		\head{Parameter}{Value}

		\entry{Learning rate $\alpha$}{0.1}
		\entry{Context look ahead $\tau$}{10~s}
		\entry{Maximum buffering time $\Delta t_{\max}$}{120~s}
		\entry{Trade-off factor $w$}{\textbf{0.8}, {0, ..., 1.0}}
		\entry{Deadline violation punishment $\Omega$}{-10}
		\entry{Connectivity map cell width $c$}{25~m}

		\bottomrule
		
	\end{tabular}
	\label{tab:parameters}
\end{table}

%
%
\newcolumntype{Y}{>{\centering\arraybackslash}X}

\begin{table}[ht]
	\centering
	\caption{\ac{MNO}-specific Paramters for \ac{RL-CAT} and \ac{RL-pCAT}}
	
	\begin{tabularx}{\columnwidth}{l c*{6}{Y}}
		\toprule
		\textbf{Model} & \multicolumn{2}{c}{\textbf{\mnoA}} & \multicolumn{2}{c}{\textbf{\mnoB}} & \multicolumn{2}{c}{\textbf{\mnoC}} \\
		\cmidrule(lr){2-3} \cmidrule(l){4-5} \cmidrule(l){6-7}
		& \textbf{UL} & \textbf{DL} & \textbf{UL} & \textbf{DL} & \textbf{UL} & \textbf{DL}\\
		\midrule

		Target data rate $S*$ & 30 & 20 & 20 & 30 & 50 & 15 \\
		Maximum data rate $S_{\max}$ & 40 & 30 & 30 & 40 & 60 & 25 \\

		\bottomrule
	\end{tabularx}
	\label{tab:mnos}
	
	\vspace{0.1cm}
	\emph{UL}: Uplink, \emph{DL}: Downlink 
\end{table}

\section{Results of the Performance Evaluation}  \label{sec:results}

In this section, the results for the simulative system optimization as well as for the real world performance evaluation are presented and discussed.

\subsection{Exploration and System Optimization} \label{sec:results_opt}

At first, we investigate the required duration for the proposed methods to converge to a satisfying performance level.
%
%
\fig{}{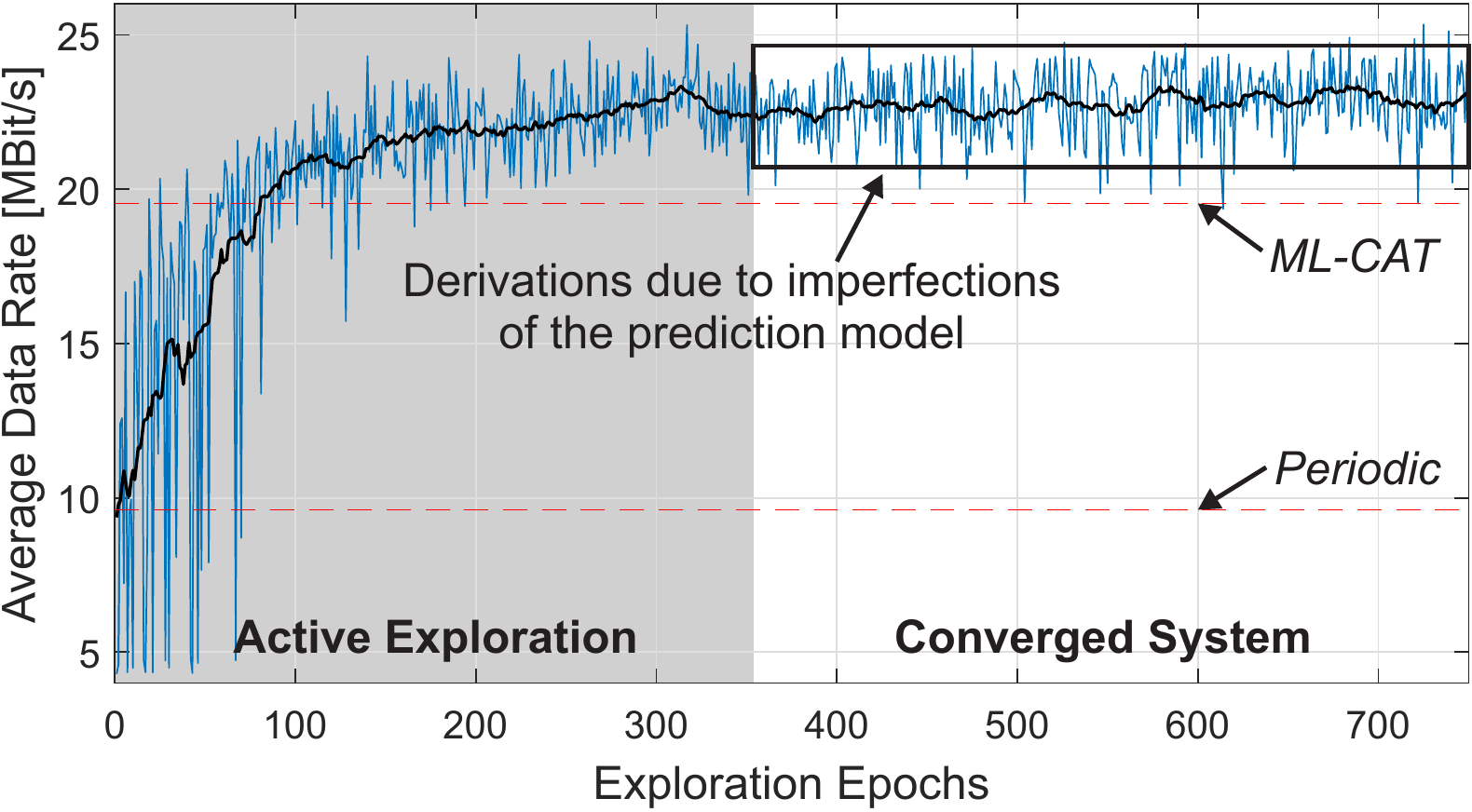}{Convergence of the virtual exploration process for uplink transmissions of \ac{MNO}~A. Each epoch corresponds to a single context trace of a \ac{DDNS} evaluation. The black line shows the average behavior.}{fig:epochs}

%
%
\begin{figure}[t]  	
	\centering		  
	\includegraphics[width=1\columnwidth]{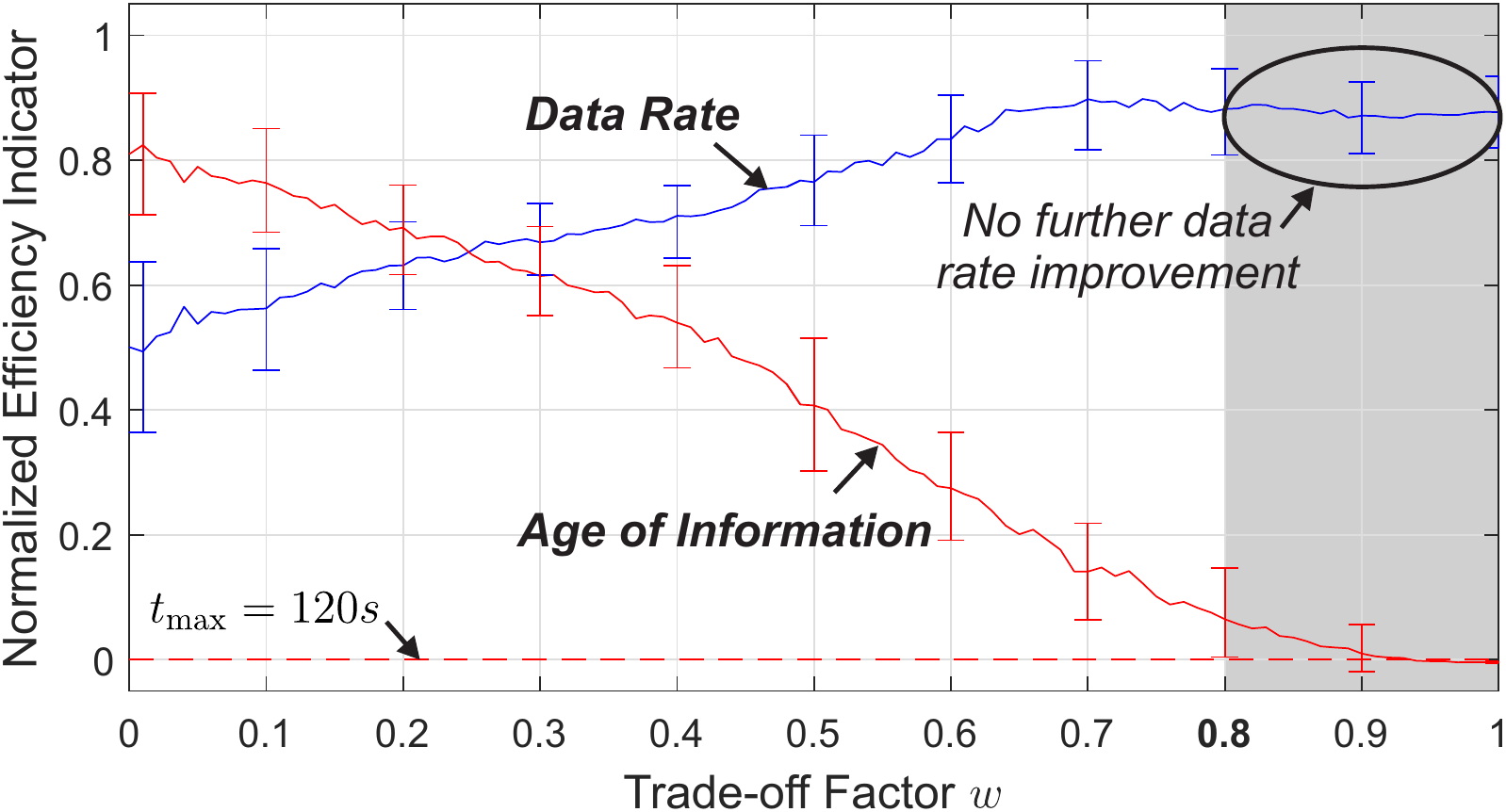}
	\caption{Trade-off between data rate efficiency $E_\text{S} = \bar{S}/S^*$ and age of information efficiency $E_\text{AoI} = 1-\bar{\Delta}t / \Delta t_{\max}$, which is controlled with the parameter $w$. The errorbar shows the standard deviation of the mean over 2400 evaluation runs per configuration.}
	\vspace{-0.5cm}	
	\label{fig:tradeoff}	
\end{figure}

%
%
\fig{}{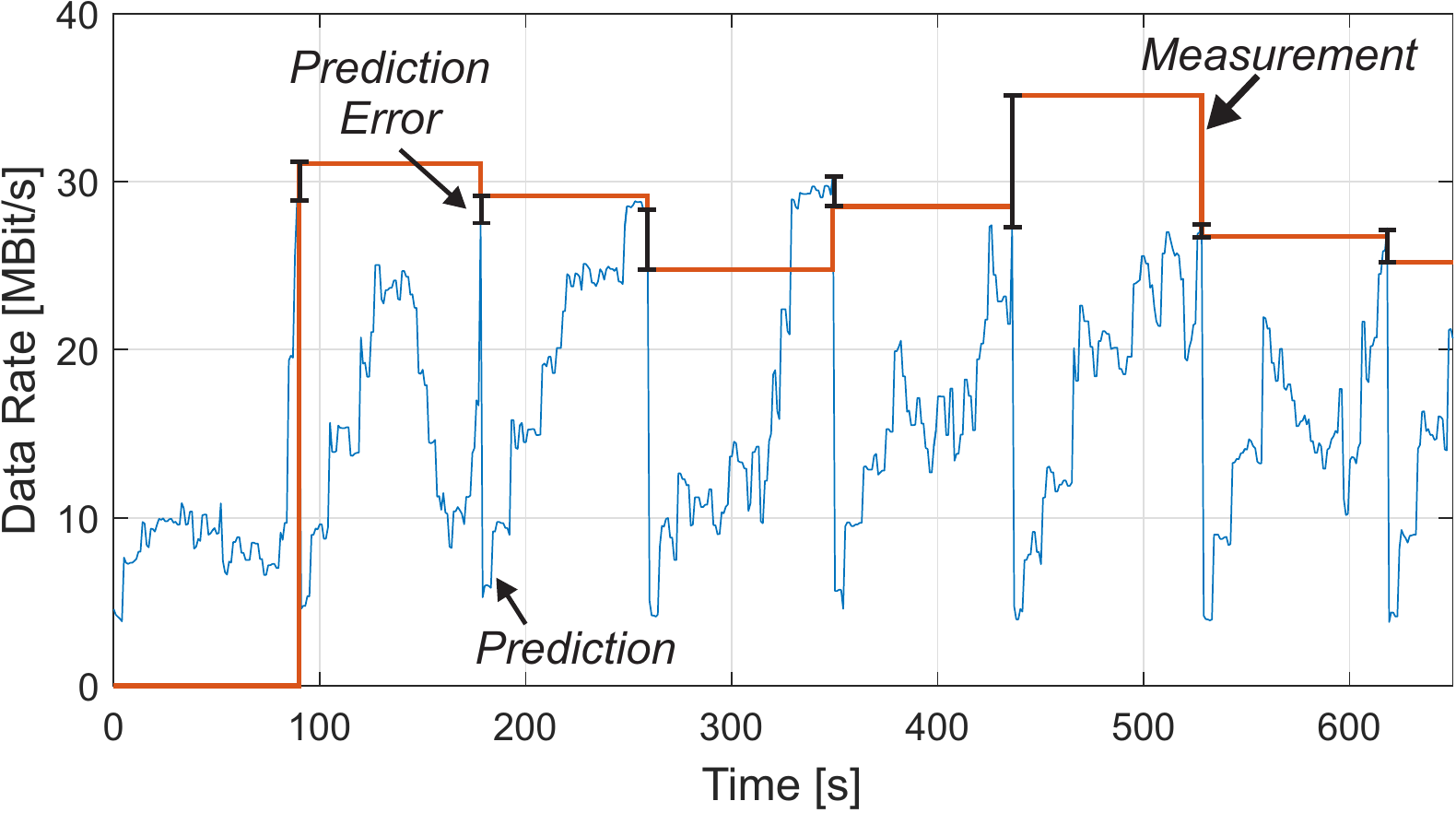}{Example temporal behavior of the \ac{RL-CAT} transmission scheme. The flanks of the measurement trace show the actual transmission times.}{fig:prediction_time}

%
%
\begin{figure}[] 
	\centering
	
	\subfig{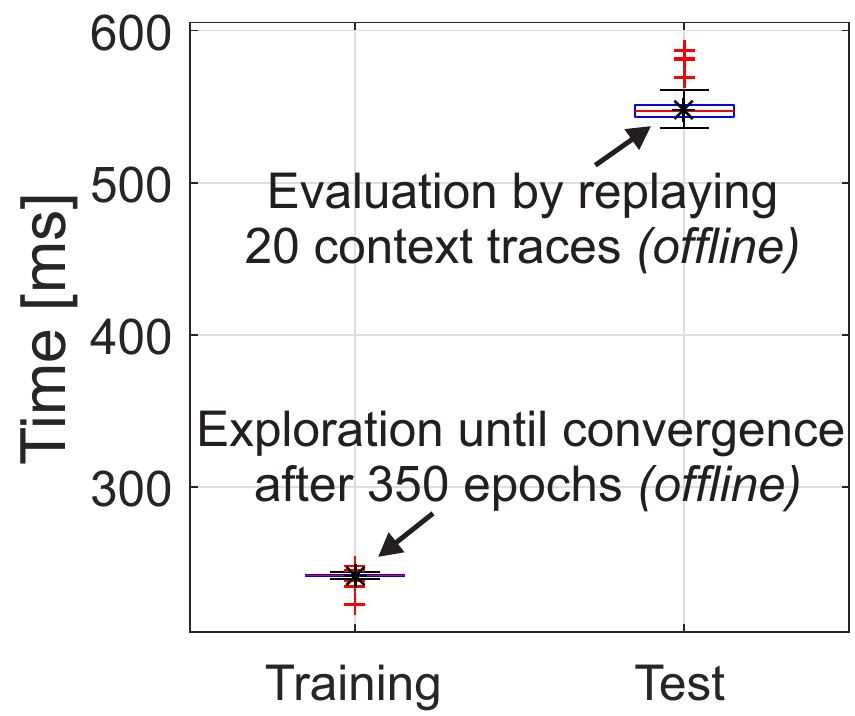}{0.24}{\ac{DDNS}-enabled Reinforcement Learning}%
	\subfig{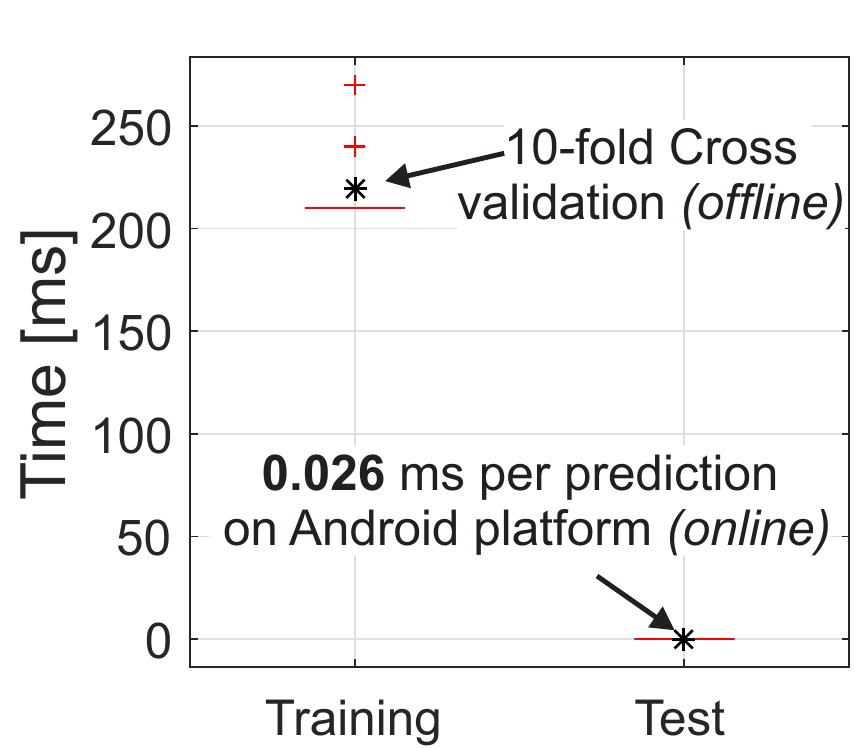}{0.24}{\ac{RF}-based Data Rate Prediction}%
	
	\caption{Temporal effort for training and applying the machine learning models.}
	\label{fig:training_times}
\end{figure}	

%
%
\begin{figure*}[t] 
	\centering
	
	\subfig{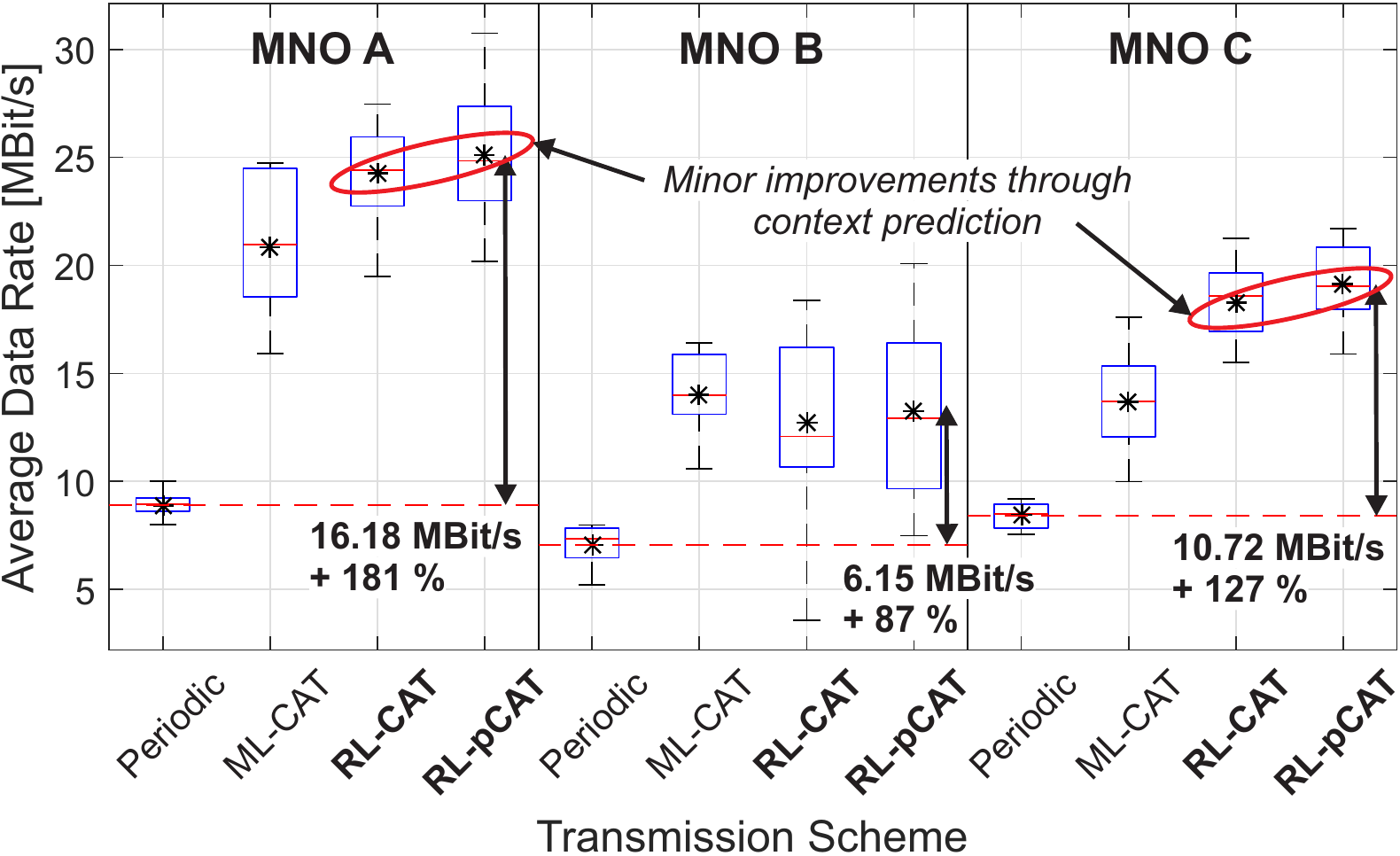}{\sfwDual}{Uplink}%
	\subfig{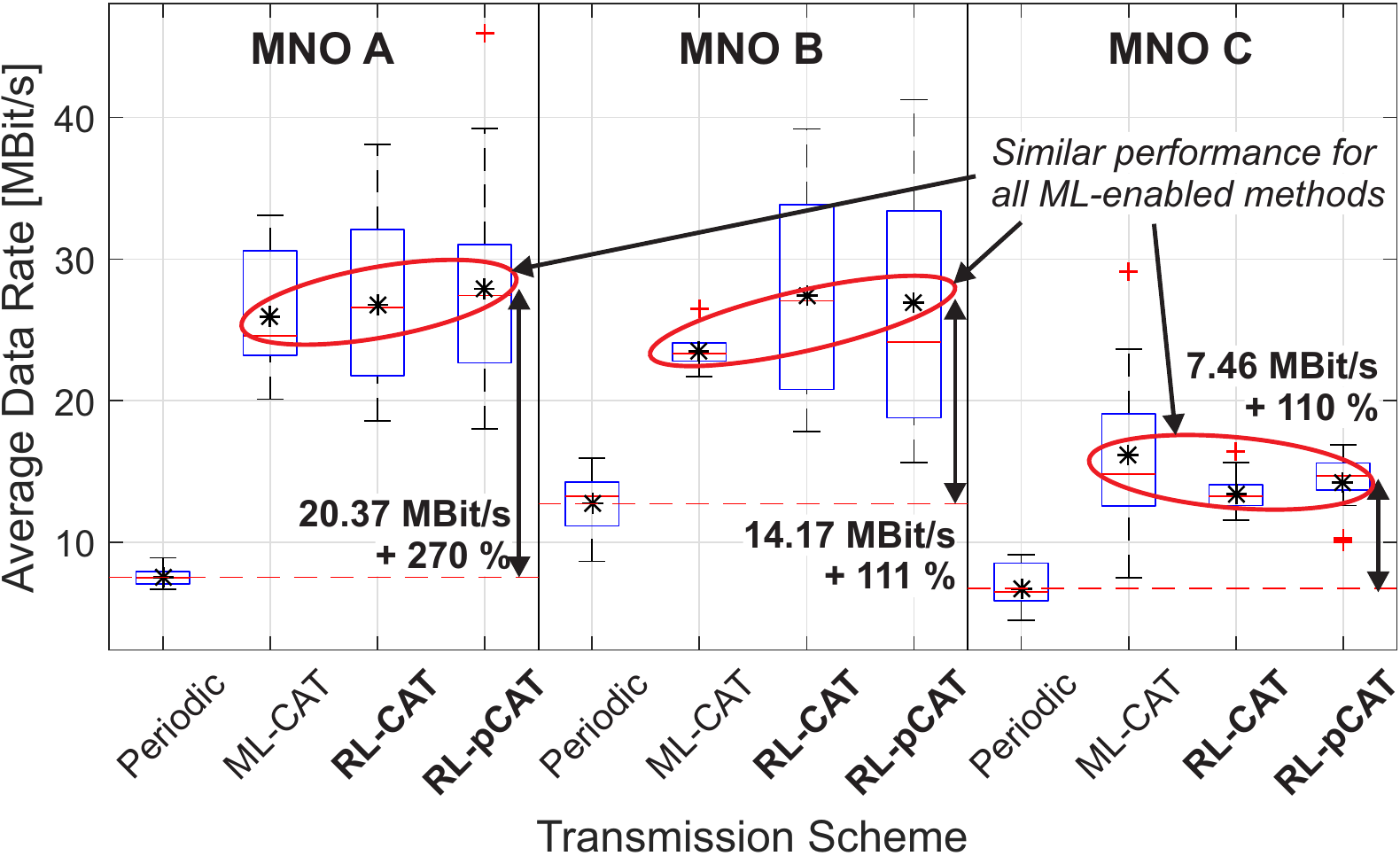}{\sfwDual}{Downlink}%

	\caption{Real world performance comparison of the resulting end-to-end data rates for the considered transmission schemes and \acp{MNO}.}
	
	\vspace{-0.5cm}
	\label{fig:boxplots_throughput}
\end{figure*}

%
%
\begin{figure*}[b] 
	\centering
	
	\vspace{-0.5cm}
	\subfig{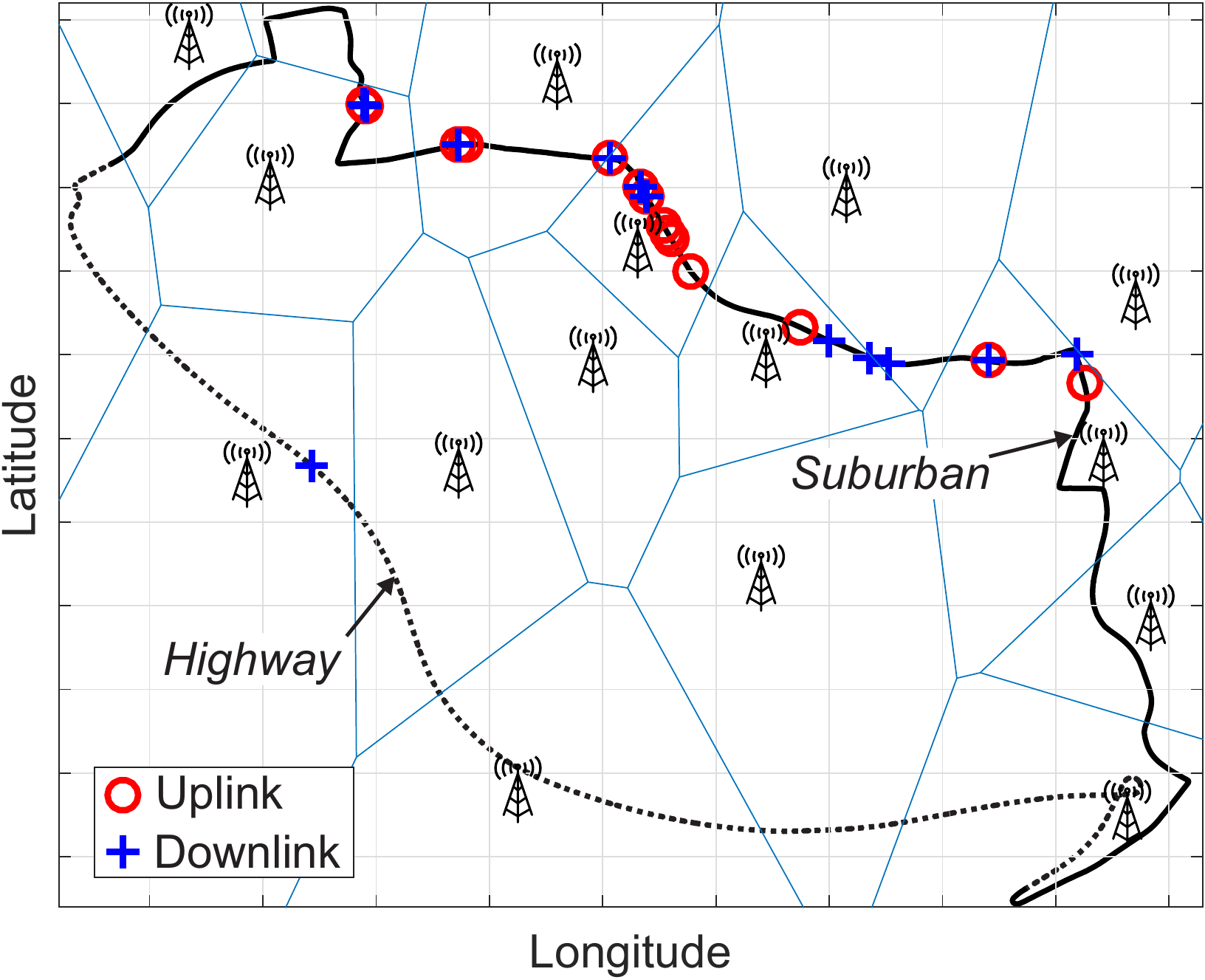}{\sfw}{\mnoA}%
	\subfig{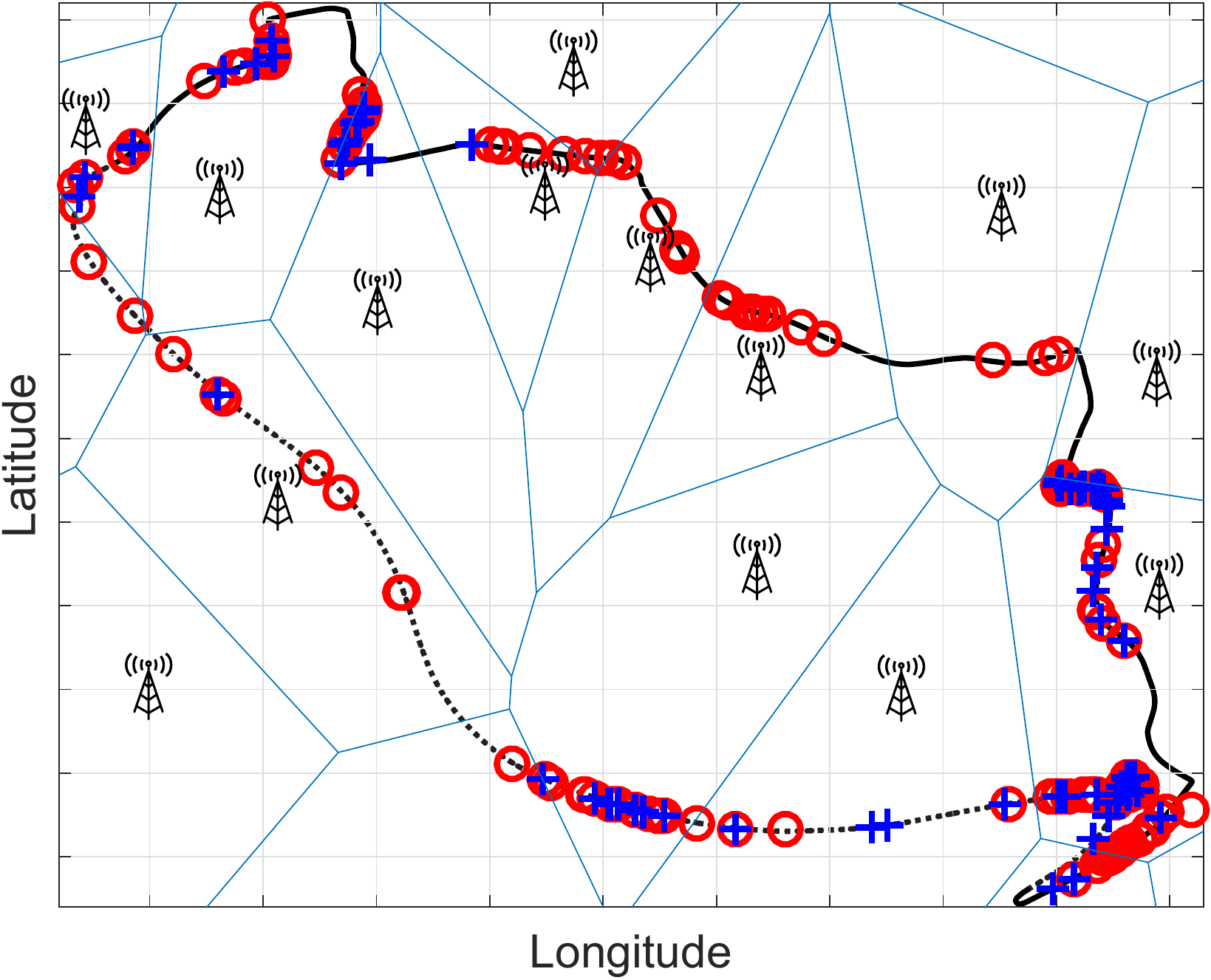}{\sfw}{\mnoB}%
	\subfig{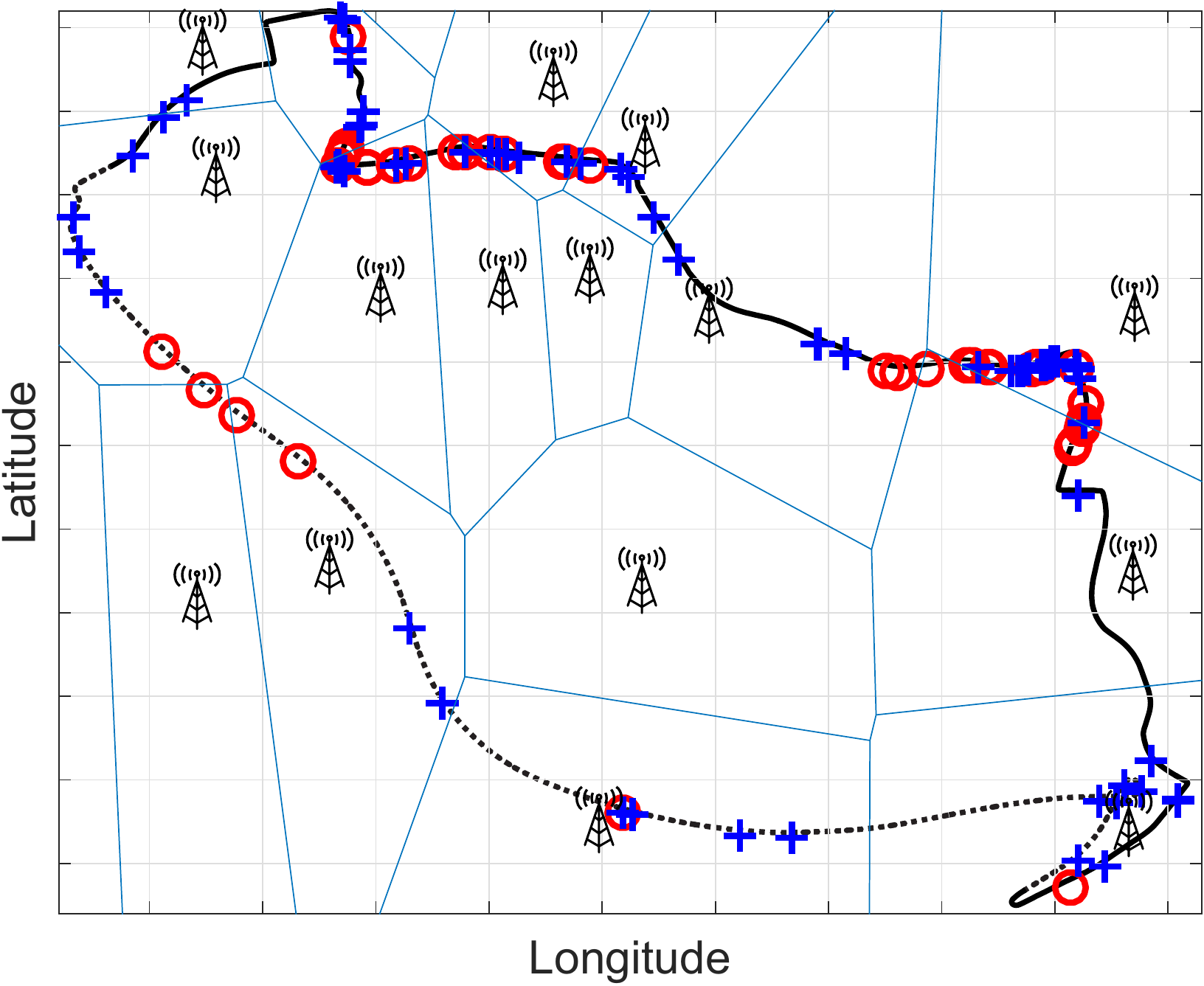}{\sfw}{\mnoC}%
	
	\caption{Geospatial distribution of the blackspot regions where the transmissions significantly deceed the average data rate of the \ac{MNO} ($S(t)<\bar{S}/2$). The voronoi diagram shows the \ac{MNO}-specific \acp{eNB} locations.}
	\label{fig:blackspot_map}
\end{figure*}	
The model convergence within the virtual exploration process is shown in Fig.~\ref{fig:epochs} for the uplink data set of \mnoA. Each epoch corresponds to a virtual single drive test synthesized by the \ac{DDNS}, which contains the time series data of the context indicators as observed by the vehicle moving on its trajectory.
%
%
For comparison, the baselines for periodic data transfer and the probabilistic \ac{ML-CAT} approach are shown. After $\sim$80 epochs, the \ac{RL-CAT} reliably outperforms the periodic approach and achieves a significant performance margin towards \ac{ML-CAT} after around $\sim$200 epochs. Convergence is reached after $\sim$350 training epochs.
%
%
It needs to be remarked that no perfect convergence can be achieved due to the remaining error range of the prediction model. Since the error-affected data rate prediction is the fundamental metric for the decision making, unfavorable decisions occur due to model imperfections.
It can also be seen that the reinforcement learning process highly benefits from the computational efficiency of the virtual exploration with \ac{DDNS}. Reaching convergence based on real world experiments would have required to perform more than 4000~km of drive tests. In contrast, \ac{DDNS} does only require $\sim$250~ms to reach convergence by learning from replaying the transmissions of the 350 context traces (Intel Core i7-4770@3.4GHz platform with 32 GB RAM).

The parameter $w$ allows to control the fundamental trade-off between data rate efficiency $E_\text{S} = \bar{S}/S^*$ and age of information efficiency $E_\text{AoI} = 1 - \bar{\Delta}t / \Delta t_{\max}$ with $\bar{S}$ being the average data rate and $\bar{\Delta}t$ being the average \ac{AoI}. For data rate optimization, the transmission scheme will rather prefer larger packet sizes in order to improve the payload-overhead ratio and to compensate the \emph{slow start} mechanism of \ac{TCP}. Fig.~\ref{fig:tradeoff} shows the impact of the $w$ for \ac{AoI} and data rate.
For $w>0.8$, the age of information exceeds $t_{\max}$ for some transmissions, which results in a negative efficiency. In the following, we apply $w=0.8$ in order to allow a performance comparison with \ac{CAT} and \ac{ML-CAT}, which focus on data rate optimization. Reduced transmission times lead to an early release of occupied resources and contribute to improving the intra-cell coexistence between multiple users \cite{Sliwa/etal/2019d}.

Fig.~\ref{fig:prediction_time} illustrates the temporal behavior of the \ac{RL-CAT} transmission scheme with respect to the predicted data rate. 
The interdependency between payload size and achievable data rate can be clearly identified. After each transmission, the transmission buffer is cleared, which results in a drop of the predicted data rate due to a reduction of the payload-overhead-ratio. With in an increased buffering time and incoming sensor data, the predicted data rate is increased again.

%
%
The temporal effort related to the training and evaluation phases of the machine learning models is shown in Fig.~\ref{fig:training_times}.

%
%
Although the data-driven exploration process is performed based on a large amount of synthesized data, it only considers the actually performed transmissions. In contrast to that, the \ac{DDNS}-based evaluation involves the replay of the whole time series data for each of the 20 context traces. Still, both parts can be processed rapidly.
%
%
Online predictions on the Android platform have a practically negligible impact on the total execution time. Since the \texttt{C++} implementation of the \ac{RF} model consists of a binary tree of \texttt{if/else} conditions, it can be evaluated in real time.

\subsection{Real World Performance Comparison}

%
%
In the following, the converged transmission schemes are applied in the real world and compared to other transmission approaches. As references, we consider straightforward periodic transmission with a fixed interval $\Delta t = 10$~s and \ac{ML-CAT}-based data transfer according to \cite{Sliwa/etal/2019d}.

The overall results of the considered transmission schemes and \acp{MNO} are illustrated in Fig.~\ref{fig:boxplots_throughput} for uplink (a) and downlink (b) direction.
%
%
As discussed in Sec.~\ref{sec:data_rate_prediction}, the data rate prediction works more accurately in the uplink as the cell is more impacted by channel-related effects than by congestion. A general observation is the that the predictive method \ac{RL-pCAT} achieves slight improvement compared to \ac{RL-CAT}, which outperforms the other approaches in most cases. In comparison to periodic data transfer, a data rate improvement by 181\% is achieved in the uplink and by 270\% in the downlink for \mnoA.
%
%
For \mnoB, the uplink data rate prediction is not very accurate ($R^2 = 0.678\pm0.04$), which leads to a slightly worse performance for \ac{RL-CAT} than for \ac{ML-CAT}. However, \ac{RL-pCAT} is able to compensate many of the outliers through its context-predictive behavior.
%
%
In the downlink transmission direction, all opportunistic approaches achieve a similar level of improvement compared to periodic transfer. For \mnoA and \mnoB, the proposed reinforcement learning-based approaches outperform the other opportunistic methods. \mnoC suffers from a low downlink data rate prediction accuracy.

%
%
Since the reinforcement-based decision making is based on the predicted data rate, future optimizations of the proposed scheme should aim to increase the prediction accuracy. A promising approach is the application of \emph{cooperative} approaches for cell load estimation. Upcoming 5G networks explicitly consider machine learning-based load analysis through a \ac{NWDAF} \cite{3GPP/2019a}. Although this method is a part of the core network, providing the acquired information for the \acp{UE} -- e.g., via the control channels -- could lead to significant improvements for client-side context-aware data transfer.

%
%
Many of the reasons for significant prediction errors are related to geospatial effects such as \emph{cellular handovers} and even technology fallbacks. In road safety management, the term \emph{blackspot} refers to regions with a high probability for road accidents. Analogously, we can define \emph{communication blackspots} which show a clustering of low data rate transmissions.

Fig.~\ref{fig:blackspot_map} provides a map of the two evaluation scenarios and the network infrastructure locations of all considered \acp{MNO}. Furthermore, all transmission that fulfill $S(t)<\bar{S}/2$ and therefore significantly deceed the average transmission performance of the \ac{MNO} are shown.
It can be seen that those transmission can be aggregated to blackspot regions for each of the \ac{MNO}. Although the voronoi diagram only considers the \ac{eNB} locations and not the real resulting coverage areas, many of the blackspot regions are close to the cell borders. 
It is very plausible that future opportunistic methods can achieve further improvements by proactively considering blackspot regions within the reinforcement learning-based transmission process.

\section{Conclusion}

%
%
In this paper, we presented a reinforcement learning-based transmission approach for optimizing the end-to-end performance of vehicular data transfer. The proposed opportunistic communication scheme schedules transmissions cognitively with respect to the predicted channel conditions.

%
%
The results of the real world performance evaluation show that the proposed approach significantly outperforms existing probabilistic channel-aware transmission schemes in most scenarios and is able to achieve massive improvements in the resulting data rate compared to typically considered periodic data transfer.

%
%
In future work, we will optimize the data rate prediction accuracy by explicitly considering blackspot context information and by applying a network-assisted load estimation approach similar to \ac{NWDAF}. On this foundation, we will furthermore develop a reinforcement learning based transmission scheme for multi-\ac{MNO} networks and investigate the performance of \emph{multi-armed bandits} and \emph{\ac{DRL}}.

\ifdoubleblind

\else
\section*{Acknowledgment}

\footnotesize
Part of the work on this paper has been supported by Deutsche Forschungsgemeinschaft (DFG) within the Collaborative Research Center SFB 876 ``Providing Information by Resource-Constrained Analysis'', project B4.
\fi

\bibliographystyle{IEEEtran}
\bibliography{Bibliography}

\begin{thebibliography}{10}
\providecommand{\url}[1]{#1}
\csname url@samestyle\endcsname
\providecommand{\newblock}{\relax}
\providecommand{\bibinfo}[2]{#2}
\providecommand{\BIBentrySTDinterwordspacing}{\spaceskip=0pt\relax}
\providecommand{\BIBentryALTinterwordstretchfactor}{4}
\providecommand{\BIBentryALTinterwordspacing}{\spaceskip=\fontdimen2\font plus
\BIBentryALTinterwordstretchfactor\fontdimen3\font minus
  \fontdimen4\font\relax}
\providecommand{\BIBforeignlanguage}[2]{{%
\expandafter\ifx\csname l@#1\endcsname\relax
\typeout{** WARNING: IEEEtran.bst: No hyphenation pattern has been}%
\typeout{** loaded for the language `#1'. Using the pattern for}%
\typeout{** the default language instead.}%
\else
\language=\csname l@#1\endcsname
\fi
#2}}
\providecommand{\BIBdecl}{\relax}
\BIBdecl

\bibitem{Bui/etal/2017a}
N.~Bui, M.~Cesana, S.~A. Hosseini, Q.~Liao, I.~Malanchini, and J.~Widmer, ``A
  survey of anticipatory mobile networking: Context-based classification,
  prediction methodologies, and optimization techniques,'' \emph{IEEE
  Communications Surveys \& Tutorials}, 2017.

\bibitem{Sliwa/etal/2018b}
B.~Sliwa, T.~Liebig, R.~Falkenberg, J.~Pillmann, and C.~Wietfeld, ``Efficient
  machine-type communication using multi-metric context-awareness for cars used
  as mobile sensors in upcoming {5G} networks,'' in \emph{2018 IEEE 87th
  Vehicular Technology Conference (VTC-Spring)}, Porto, Portugal, Jun 2018,
  {Best Student Paper Award}.

\bibitem{Sliwa/etal/2019d}
B.~Sliwa, R.~Falkenberg, T.~Liebig, N.~Piatkowski, and C.~Wietfeld, ``Boosting
  vehicle-to-cloud communication by machine learning-enabled context
  prediction,'' \emph{IEEE Transactions on Intelligent Transportation Systems},
  Jul 2019.

\bibitem{Sliwa/Wietfeld/2019b}
B.~Sliwa and C.~Wietfeld, ``Empirical analysis of client-based network quality
  prediction in vehicular multi-{MNO} networks,'' in \emph{2019 IEEE 90th
  Vehicular Technology Conference (VTC-Fall)}, Honolulu, Hawaii, USA, Sep 2019.

\bibitem{Sliwa/2019a}
\BIBentryALTinterwordspacing
B.~Sliwa, ``Raw data of real world measurements,'' Oct 2019. [Online].
  Available: \url{https://doi.org/10.5281/zenodo.3490335}
\BIBentrySTDinterwordspacing

\bibitem{Jiang/etal/2017a}
C.~Jiang, H.~Zhang, Y.~Ren, Z.~Han, K.~C. Chen, and L.~Hanzo, ``Machine
  learning paradigms for next-generation wireless networks,'' \emph{IEEE
  Wireless Communications}, vol.~24, no.~2, pp. 98--105, April 2017.

\bibitem{Ye/etal/2018a}
H.~Ye, L.~Liang, G.~Y. Li, J.~Kim, L.~Lu, and M.~Wu, ``Machine learning for
  vehicular networks: {R}ecent advances and application examples,'' \emph{IEEE
  Vehicular Technology Magazine}, vol.~13, no.~2, pp. 94--101, June 2018.

\bibitem{Sutton/Barto/2018a}
R.~S. Sutton and A.~G. Barto, \emph{Reinforcement learning: {A}n introduction},
  2nd~ed.\hskip 1em plus 0.5em minus 0.4em\relax The MIT Press, 2018.

\bibitem{Gacanin/2019a}
H.~{Gacanin}, ``Autonomous wireless systems with artificial intelligence: {A}
  knowledge management perspective,'' \emph{IEEE Vehicular Technology
  Magazine}, pp. 1--1, 2019.

\bibitem{Mangla/etal/2016a}
T.~Mangla, N.~Theera-Ampornpunt, M.~Ammar, E.~Zegura, and S.~Bagchi, ``Video
  through a crystal ball: {E}ffect of bandwidth prediction quality on adaptive
  streaming in mobile environments,'' in \emph{Proceedings of the 8th
  International Workshop on Mobile Video}, ser. MoVid '16.\hskip 1em plus 0.5em
  minus 0.4em\relax New York, NY, USA: ACM, 2016, pp. 1:1--1:6.

\bibitem{Bouali/etal/2016a}
F.~{Bouali}, K.~{Moessner}, and M.~{Fitch}, ``A context-aware user-driven
  framework for network selection in {5G} multi-{RAT} environments,'' in
  \emph{2016 IEEE 84th Vehicular Technology Conference (VTC-Fall)}, Sep. 2016,
  pp. 1--7.

\bibitem{Akselrod/etal/2017a}
M.~Akselrod, N.~Becker, M.~Fidler, and R.~Luebben, ``{4G} {LTE} on the road -
  what impacts download speeds most?'' in \emph{2017 IEEE 86th Vehicular
  Technology Conference (VTC-Fall)}, Sep. 2017, pp. 1--6.

\bibitem{Riihijarvi/Mahonen/2018a}
J.~Riihijarvi and P.~Mahonen, ``Machine learning for performance prediction in
  mobile cellular networks,'' \emph{IEEE Computational Intelligence Magazine},
  vol.~13, no.~1, pp. 51--60, Feb 2018.

\bibitem{Jomrich/etal/2018a}
F.~Jomrich, A.~Herzberger, T.~Meuser, B.~Richerzhagen, R.~Steinmetz, and
  C.~Wille, ``Cellular bandwidth prediction for highly automated driving -
  {E}valuation of machine learning approaches based on real-world data,'' in
  \emph{Proceedings of the 4th International Conference on Vehicle Technology
  and Intelligent Transport Systems 2018}, no.~4.\hskip 1em plus 0.5em minus
  0.4em\relax SCITEPRESS, Mar 2018, pp. 121--131.

\bibitem{Samba/etal/2017a}
A.~Samba, Y.~Busnel, A.~Blanc, P.~Dooze, and G.~Simon, ``Instantaneous
  throughput prediction in cellular networks: {W}hich information is needed?''
  in \emph{2017 IFIP/IEEE Symposium on Integrated Network and Service
  Management (IM)}, May 2017, pp. 624--627.

\bibitem{Breiman/2001a}
L.~Breiman, ``Random forests,'' \emph{Mach. Learn.}, vol.~45, no.~1, pp. 5--32,
  Oct. 2001.

\bibitem{LeCun/etal/2015a}
Y.~LeCun, Y.~Bengio, and G.~Hinton, ``\BIBforeignlanguage{English (US)}{Deep
  learning},'' \emph{\BIBforeignlanguage{English (US)}{Nature}}, vol. 521, no.
  7553, pp. 436--444, 5 2015.

\bibitem{Sliwa/Wietfeld/2019c}
B.~Sliwa and C.~Wietfeld, ``Towards data-driven simulation of end-to-end
  network performance indicators,'' in \emph{2019 IEEE 90th Vehicular
  Technology Conference (VTC-Fall)}, Honolulu, Hawaii, USA, Sep 2019.

\bibitem{Cavalcanti/etal/2018a}
E.~R. Cavalcanti, J.~A.~R. de~Souza, M.~A. Spohn, R.~C. d.~M. Gomes, and A.~F.
  B. F.~d. Costa, ``{VANETs}' research over the past decade: {O}verview,
  credibility, and trends,'' \emph{SIGCOMM Comput. Commun. Rev.}, vol.~48,
  no.~2, pp. 31--39, May 2018.

\bibitem{Lee/etal/2019a}
J.~Lee, J.~Lee, Y.~Im, S.~Dhawaskar~Sathyanarayana, P.~Rahimzadeh, X.~Zhang,
  M.~Hollingsworth, C.~Joe-Wong, D.~Grunwald, and S.~Ha, ``{CASTLE} over the
  air: {D}istributed scheduling for cellular data transmissions,'' in
  \emph{Proceedings of the 17th Annual International Conference on Mobile
  Systems, Applications, and Services}, ser. MobiSys '19.\hskip 1em plus 0.5em
  minus 0.4em\relax New York, NY, USA: ACM, 2019, pp. 417--429.

\bibitem{Watkins/Dayan/1992a}
C.~J. C.~H. Watkins and P.~Dayan, ``Q-learning,'' \emph{Machine Learning},
  vol.~8, no.~3, pp. 279--292, May 1992.

\bibitem{Sliwa/etal/2018a}
B.~Sliwa, T.~Liebig, R.~Falkenberg, J.~Pillmann, and C.~Wietfeld, ``Machine
  learning based context-predictive car-to-cloud communication using
  multi-layer connectivity maps for upcoming {5G} networks,'' in \emph{2018
  IEEE 88th Vehicular Technology Conference (VTC-Fall)}, Chicago, USA, Aug
  2018.

\bibitem{Hall/etal/2009a}
M.~Hall, E.~Frank, G.~Holmes, B.~Pfahringer, P.~Reutemann, and I.~H. Witten,
  ``The {WEKA} data mining software: {A}n update,'' \emph{SIGKDD Explorations},
  vol.~11, no.~1, pp. 10--18, 2009.

\bibitem{3GPP/2019a}
3GPP, ``{5G System; Network Data Analytics Services;Stage 3},'' {3rd Generation
  Partnership Project (3GPP)}, Technical Specification (TS) 29.520, Mar 2019,
  version 15.3.0.

\end{thebibliography}

\end{document}